\documentclass[sigconf,natbib=false]{acmart}

\AtBeginDocument{%
  }

\setcopyright{acmlicensed}
\copyrightyear{2024}
\acmYear{2024}

\acmConference[arXiv]{Make sure to enter the correct
  conference title from your rights confirmation emai}{October,
  2024}{}

\RequirePackage[
  datamodel=acmdatamodel,
  style=acmnumeric,
  ]{biblatex}

\begin{document}

\title[Cyber-physical and business perspectives using FDT in multinational and multimodal transportation systems]{Cyber-physical and business perspectives using Federated Digital Twins in multinational and multimodal transportation systems}

\author{Ricardo M. Czekster}
\authornote{All authors contributed equally to this research.}
\authornotemark[1]
\affiliation{%
  \institution{School of Computer Science \& DT\\Aston University}
  \city{Birmingham}
  \country{UK}
}
\email{r.meloczekster@aston.ac.uk}

\author{Alexeis Garcia Perez}
\affiliation{%
  \institution{Aston Business School\\Aston University}
  \city{Birmingham}
  \country{UK}
}
\email{a.garciaperez@aston.ac.uk}

\author{Manolya Kavakli-Thorne}
\affiliation{%
  \institution{Aston Digital Futures Institute\\Aston University}
  \city{Birmingham}
  \country{UK}
}
\email{m.kavakli-thorne@aston.ac.uk}

\author{Seif Allah El Mesloul Nasri}
\affiliation{%
 \institution{Aston Digital Futures Institute\\Aston University}
 \city{Birmingham}
 \country{UK}
}
\email{s.nasri@aston.ac.uk}

\author{Siraj Shaikh}
\affiliation{%
 \institution{School of Math. \& Computer Science\\Swansea University}
 \city{Swansea}
 \country{UK}
}
\email{s.a.shaikh@swansea.ac.uk}

\renewcommand{\shortauthors}{Czekster et al.}

\begin{abstract}
Digital Twin (DT) technologies promise to remove cyber-physical barriers in systems and services and provide seamless management of distributed resources effectively.
Ideally, full-fledged instantiations of DT offer bi-directional features for physical-virtual representations, tackling data governance, risk assessment, security and privacy protections, resilience, and performance, to name a few characteristics.
More broadly, Federated Digital Twins (FDT) are distributed physical-virtual counterparts that collaborate for enacting synchronisation and accurate mapping of multiple DT instances.
In this work we focus on understanding and conceptualising the cyber-physical and business perspectives using FDT in multinational and multimodal transportation systems.
These settings enforce a plethora of regulations, compliance, standards in the physical counterpart that must be carefully considered in the virtual mirroring.
Our aim is to discuss the regulatory and technical underpinnings and, consequently, the existing operational and budgetary overheads to factor in when designing or operating FDT.
\end{abstract}



\received{11 Oct 2024}
\received[revised]{11 Oct 2024}
\received[accepted]{11 Oct 2024}

\maketitle

\section{Introduction}

Multinational and multimodal transportation systems (MMTS) concern investigating the movement of goods and people over road, rail, water, and air, aiming seamless coordination.
Its major stakeholders are for instance motorists, transit-riders, freight operators, pedestrians, and pilots, to name a few.
Problems related to these settings consider planning, coordination, congestion control, synchronisation, and systems integration.
Given the scale of the issues behind MMTS new technologies have emerged in the past years to tackle these issues through effective remote management whilst allowing predictive analytics in (near) real-time.

Among these new technologies, adopting Digital Twins (DT)~\cite{barricelli2019survey,mihai2022digital} to remotely observe the system has been the centre of attention in recent years~\cite{vanderhorn2021digital,semeraro2021digital}.
In an attempt to understand the proposition, Mihai et al. (2022)~\cite{mihai2022digital} listed five competing `definitions' of DT.
In this work, we adopt these authors' description, where DT consist of a self-adapting, self-regulating, self-monitoring, and self-diagnosing system-of-systems (SoS) that has a clear link between its physical and virtual counterparts representing its composing systems with fidelity and adequate level of synchronisation where enabling technologies support its objectives and services by adding operational and business value to the physical entity.
Note that in this context, physical objects in DT are for instance people, objects, spaces, systems, and processes.
Recent research has described efforts to characterise DT in a host of scientific literature, outlining enabling technologies, standards, reference architecture models, open challenges, and multiple definitions adapted to the context where they sit~\cite{fuller2020digital,jones2020characterising,ferko2023standardisation,hakiri2024comprehensive}.

We refer to DT as a physical representation reality that plays an important part in the analysis, not only used as a `model' of a setting in the virtual counterpart only~\cite{wright2020tell,chang2024digital}.
The DT provision continuously collects real-time data from the physical counterpart using a range of sensing and tracking technologies to provide an accurate mapping of the current state of the SoS ensemble at a given point~\cite{yun2017data}.
It fosters comprehensive situational awareness allowing it to devise predictive analytics models of potential future states, accommodating operational constraints and enacting reasonable asset allocation.
More recently, special attention has been given to cyber-physical security issues in DT~\cite{pokhrel2020digital} in approaches known as Security Enhancing DT (SEDT), with practical applications when hardening systems and services against malicious threat agents.

Federated Digital Twins (FDT)~\cite{vergara2023federated,jeong2022digital,broo2022design,yang2021developments} have been identified as a potential solution to establish communication among multiple DT, to achieve cooperation in a virtual space, based on the representation of large and complex systems to assist in the optimisation and efficiency in the entire life cycle of products and services~\cite{jeong2022digital}.
However, FDT's characteristics, benefits, and limitations have not been widely investigated, and coordination and synchronisation mechanisms between DT in a federated environment have not been addressed. 
There are some advances in the description of these elements in recent literature, for example, 
Yun et al. (2017)~\cite{yun2017data} devised a framework for coordinating distributed digital twin cooperation.
FDT pose new challenges for stakeholders, as it not only synchronises activities, but it should also cater for identifying cooperation and management, handle metadata exchanges and underlying intelligent technologies, mutual information, and governance~\cite{jeong2022digital}.
The contribution here is two-fold:
\begin{enumerate}
   \item Inspect the state-of-the-art of DT/FDT, selecting case studies, tools, and frameworks, and commenting on general principles for effectively deploying them in operational contexts of multinational and multimodal transportation settings.
   \item Discuss standardisation, data governance, compliance, and regulatory framework for FDT in multinational scope\footnote{We use these concepts interchangeably in this work. Existing literature differentiates them with regards to either centralised or decentralised management, respectively.}, focusing on integration, synchronisation, and management outlining cyber-physical and business perspectives.
\end{enumerate}

The goal of this paper is to investigate \textit{``how to operationalise effective FDT seamlessly, closely reflecting cyber-physical interactions in multinational and multimodal transportation systems?''}

The paper is organised as follows.
Section~\ref{s:related} discusses related work and contextualisation and Section~\ref{s:fdt-cp-business} presents our discussion on FDT in MMTS.
In Section~\ref{s:conc} we outline final considerations.

\section{Context and related work}\label{s:related}
The literature of DT is vast and comprehensive as research has been quite fruitful in recent years.
Grieves (2023)~\cite{grieves2023digital} has outlined and discussed the past, present, and future of DT in the wake of supporting new technologies that aggregate new value to the proposition (eg, 5G/6G networks).
Recently, Hakiri et al. (2024)~\cite{hakiri2024comprehensive} discussed standardisation, open challenges, and latest enabling technologies and DT use cases in a survey.
Broo et al. (2022)~\cite{broo2022design} discussed DT design and implementation in smart infrastructure, outlining systems, information and organisational perspectives and how they interplay for providing a wide range of features to stakeholders.

Chang et al. (2024)~\cite{chang2024digital} investigated key enabling technologies for DT in transportation systems whereas Gao et al. (2021)~\cite{gao2021digital} discussed the use of DT in transportation infrastructure focusing on DT in transportation and assessing four areas namely railway, highway, bridge, and tunnel.
DT finds useful applications in transportation infrastructure given the sheer scale of the problem and the potential opportunities it unveils to stakeholders, eg, Cress et al. (2021)~\cite{cress2021intelligent} investigated Intelligent Transportation Systems (ITS).
Authors highlighted further required research on effectively using sensors to map the physical reality, incorporation of plug-and-play mechanisms, and how to achieve secure real-time distribution of the ensemble of DT in a decentralised fashion.
ITS has been used with DT for improving transportation systems~\cite{figueiredo2001towards,alam2016introduction,bao2021review}.
Chaalal et al. (2023)~\cite{chaalal2023integrating} discussed mobility issues for integrating connected and automated shuttles for transporting passengers and goods.

With respect to novel technologies acting as enablers of the DT vision, we mention AI/ML~\cite{rathore2021role,barricelli2019survey,minerva2023artificial,bariah2024interplay}, blockchains~\cite{yaqoob2020blockchain}, augmented reality and visualisation for the simulation of what-if scenarios~\cite{schroeder2016visualising}, and novel IoT for better tracking and safety concerns~\cite{zhao2021iot}, to mention a few modern approaches.
The direction taken by Nguyen et al. (2024)~\cite{nguyen2024security} was to incorporate AI/ML into DT and commenting on a framework for \textit{explainability} and security services orchestration.

\section{FDT: Cyber-physical \& business issues}\label{s:fdt-cp-business}
Successful deployment of DT in complex real-world cases require collaboration among individual systems.
The federation generates a virtual medium for data exchange between the DT.
FDT allow interconnection among autonomous DT in the virtual space, as a network of cooperative DT to assist the overall operations through knowledge acquisition and reasoning, leading to an informed and intelligent decision making platform.
There are various architectural styles for data governance for FDT.
They are categorised as Centralised, Peer-to-peer, Hierarchical, and Regional~\cite{vergara2023federated,jeong2022digital}.

\begin{figure}[!htb]
  \centering
  \includegraphics[width=\linewidth]{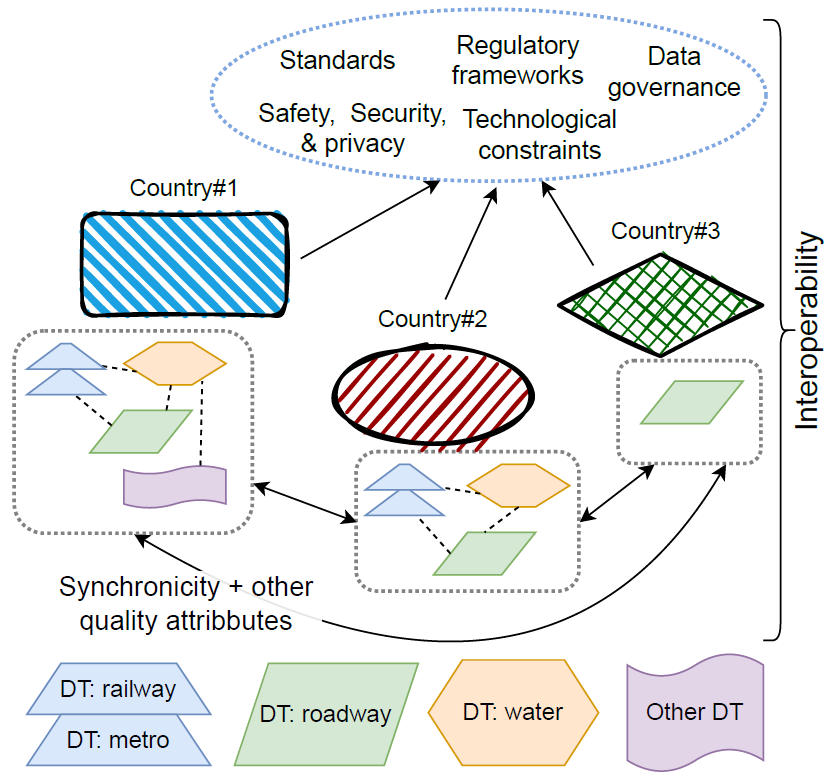}
  \caption{Overview of FDT in multinational and MMTS.}
  \Description{Boxes representing countries and the Federated Digital Twins setting.}
  \label{fig:overview-fdt}
\end{figure}

FDT require multi-layered approaches for interoperability.
Figure~\ref{fig:overview-fdt} depicts a general FDT proposition and outlines the required interaction and heterogeneity of underlying systems to provide seamless services to operators, managers, and users.

Jeong et al. (2022)~\cite{jeong2022digital} also discuss transitioning to a fully autonomous FDT state where approaches traverse five stages, namely: 1. Mirroring (replicating physical reality into DT; 2. Monitoring (tracking objects and persisting data); 3. Analytics (using Modelling \& Simulation or other analysis technique for predictive analysis or decision making); 4. Federation (for inter-operations with other DT whilst interacting with physical objects); and 5. Autonomous optimisation (for understanding, recognising, and autonomously optimising and reconfiguring the DT for better settings).

\noindent
\textbf{\textit{Business perspective:}} DT/FDT have the potential to emphasise cost savings, regulatory compliance, risk management, stakeholder collaboration, strategic planning and sustainability.
By focusing on these areas, businesses can leverage FDT to enhance operations, achieve competitive advantages and create more efficient, safe and sustainable transportation networks.

The deployment of FDT offers substantial business value and ROI by optimising operations through predictive maintenance and efficient resource allocation, leading to significant cost savings~\cite{lim2020state}.
Real-time insights provided by FDT enable better decision-making for logistics, fleet management, and operational strategies.
Moreover, the integration of FDT opens opportunities for new business models, eg, offering predictive maintenance services, dynamic routing solutions, and real-time supply chain visibility.

\noindent
\textbf{\textit{Compliance to standards:}} Compliance with national and international regulations on data governance, privacy and security is crucial.
Standardisation across FDT implementations facilitates smoother operations and interoperability, adhering to frameworks provided by ISO/UNECE.
Additionally, early risk identification through FDT allows businesses to develop mitigation strategies, ensuring smoother operations and reducing disruptions.

Specific to DT/FDT, ISO/IEC 30173:2023~\cite{iso30173:2023} mentions concepts and terminology for DT whereas IEEE 1451 defines standards for operating smart sensor digital twin federation for IoT and CPS research~\cite{lee2000ieee,song2019ieee}.
ISO 23247 proposed a standard comprising of a reference architecture for DT in manufacturing~\cite{ferko2023standardisation} and ISO 14813-1:2015 discusses reference model architectures for ITS~\cite{iso14813:2015}.

DT encompasses not only technical cyber-physical concerns but it has also repercussions over surrounding issues in business and regulations.
In terms of risk management and assessment we mention ISO 31000:2018~\cite{iso31000:2018} and its accompanying ISO 31010:2019~\cite{iso31010:2019} for tackling assessment techniques.
Implementing robust cybersecurity measures within FDT protects sensitive data, guided by frameworks such as ISO 27001:2022~\cite{iso27001:2022}.

The automotive industry has its own set of standards and regulations, provided by ISO 26262 that addresses functional safety of road vehicles~\cite{iso26262:2011}.
NIST (US) has produced a standard draft that offers a definition of DT as well as features, functions, and operational uses of surrounding technologies~\cite{voas2021considerations}.
On the other hand, ENISA (EU) discusses a holistic approach towards Critical Infrastructure\footnote{ENISA Critical Infrastructure website. Link: \url{https://www.enisa.europa.eu/topics/critical-information-infrastructures-and-services}.}.

\subsection{Multinational and Multimodal DT}
The multinational nature of multimodal transportation systems involves a host of stakeholders, regulations, technologies, and standards which add to the complexity of a FDT. This necessitates a multi-layered approach to consider national regulations over international ones.  
In order to accommodate all these constraints, the FDT proposition must consider existing trade-offs among different quality attributes without sacrificing informative outcomes to stakeholders.
FDT can be considered as a meta system referring to the concept of system-of- systems. There could exist a situation where the system-of-systems is so constrained when enforcing a set of these quality attributes that it prevents it from generating valuable information for decision making.
One could for instance factor in requirements for scalability, interoperability, synchronicity (ie, ability to coincide physical and virtual counterparts), connectivity, performance, adaptability (ability to change), resilience (sustain operation under duress), as well as security, safety, and privacy.
Every choice is between convenience and value-added prospects, in light of technological constraints.

MMTS are complex due to their involvement in multiple countries and various transport modes like road, rail, air, and water.
As mentioned, DT and FDT technologies offers promising solutions for managing these systems providing real-time digital replicas of physical assets, processes, and systems, thus allowing better monitoring and optimisation of transportation networks. 

A primary challenge in implementing DT/FDT in MMTS is integrating diverse regulatory frameworks across countries.
Each country has its own transportation, safety, environmental, and operational standards/protocols. 
Ensuring synchronisation and interoperability across these frameworks is crucial for MMTS effectiveness, requiring standardised protocols for real-time data exchange. 

Compliance with regulatory frameworks in different countries requires various types of data.
Standardised data protocols ensure interoperability across transportation systems and regions.
For instance, the United Nations Economic Commission for Europe (UNECE) has developed standards for the digitisation of intermodal transport data exchanges to support global supply chain cargo movements, all aligned under the UN/CEFACT Multimodal Transport Reference Data Model (MMT RDM)~\cite{unece2017}.

Real-time traffic and mobility data help comply with regulatory requirements related to traffic management and safety~\cite{dasgupta2021transportation}.
Regulatory data includes information on legal requirements for transportation in each country, such as safety regulations and environmental standards~\cite{budiardjo2021digital}.
Operational data provides real-time information on transportation operations, helping manage activities across borders~\cite{dasgupta2021transportation}.
Environmental data involves emissions and fuel consumption regulated differently across countries~\cite{jamil2022comprehensive}.
Safety data includes information on transportation safety performance, such as accident reports and maintenance records.
Technological data covers information on the technologies used in transportation systems, such as interoperability standards~\cite{guerrero2015integration,wang2021multi}.

\begin{figure}[!htb]
  \centering
  \includegraphics[width=\linewidth]{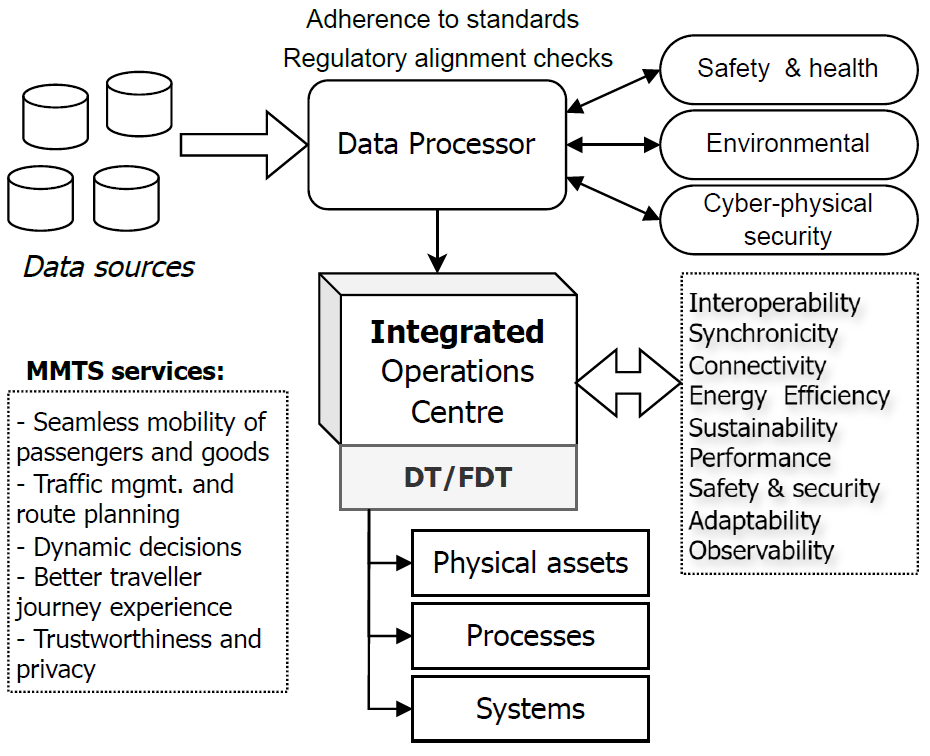}
  \caption{General functions of an Integrated Operations Centre for enabling DT/FDT services.}
  \Description{Boxes representing functions for an Integrated Operations Centre for DT/FDT.}
  \label{fig:statement}
\end{figure}

Figure~\ref{fig:statement} depicts functions of an \textit{Integrated Operations Centre} for handling and processing multiple data sources and enacting a comprehensive \textit{situational awareness} solution for DT/FDT managers.
The ``Data Processor'' activity should not only validate data and check compliance, but also perform compilation, deduplication (removing duplicates), aggregating, and prioritising entries for timely consumption in the `Integrated Operations Centre'.

Primary modes of transport for international movement include air, maritime, rail, and road transport, each presenting unique challenges and opportunities for DT, eg, road transport requires real-time traffic management and route optimisation.
It enables dynamic adjustments to traffic flows and reduce congestion by providing real-time data on vehicle status, traffic conditions, and route planning~\cite{dasgupta2021transportation}.
Rail transport involves precise scheduling and maintenance data.
They help by offering detailed insights into the operational status of trains and rail infrastructure, facilitating better scheduling and predictive maintenance to avoid disruptions~\cite{chang2024digital}.

Air transport needs detailed monitoring of flight operations and air traffic control.
Implementing DT in aviation can optimise flight operations, enhance safety through better predictive maintenance, and streamline air traffic control by simulating and analysing flight paths in real-time~\cite{jamil2022comprehensive}.
Maritime transport requires real-time tracking and environmental monitoring for ships and port operations.
DT in maritime transport optimises cargo handling, real-time tracking of vessel movements, and environmental monitoring to ensure compliance with regulations~\cite{wang2021multi}.
Integrating DT/FDT in MMTS through scalable and interoperable solutions is key for optimising transportation networks management and ensuring smooth operation across different countries and transportation modes ~\cite{budiardjo2021digital}.
This requires comprehensive data management, regulatory alignment, and technological integration, emphasising the need for ongoing research and development~\cite{ivanov2021digital,ambra2020agent}.
By easing data exchange and standardisation across different modes of transport and regulatory environments, DT/FDT can help overcome the challenges posed by regulatory diversity and technological integration ~\cite{guerrero2015integration,zheng2019application}.

Accommodating the diverse regulatory requirements, integrating real-time data across MMTS, and ensuring interoperability and security with DT/FDT can significantly enhance the efficiency, safety, and sustainability.
Achieving these benefits, however, requires coordinated efforts across stakeholders and sectors.

\subsection{Going forward: Federated DT}
The modern vision is transitioning to a cyber-physical setting that integrates multiple CPS that synchronises activities by aggregating data inputs from a host of diverse DT.
Kanimozhi et al. (2024)~\cite{kanimozhi2024intelligent} mentions the integration of DT within Smart Cities and Smart Grid whilst employing and integrating with ITS.
The authors advocate a holistic approach for tackling a myriad of systems helping societal interactions for coordinating and improving a city's transit, energy provision, and timely data analytics. 

Deploying DT/FDT poses challenges for public administration in terms of regulatory frameworks, a topic investigated by Bundin et al. (2021)~\cite{bundin2021legal}.
The authors discuss legal regulations for addressing information security, privacy, and liability issues arising in these settings.
There are discussions on using DT in supply chains to ensure smooth distribution and mobility of passengers and goods using digitalisation~\cite{marmolejo2020design}.

On the topic of business DT, Lim et al. (2020)~\cite{lim2020state} investigated product lifecycle management and business innovations in a survey.
The authors identified eight perspectives, namely, 1. Modular DT; 2. Modelling consistency and accuracy; 3. aggregating Big Data analytics; 4. Simulation improvements; 5. Virtual Reality integration; 6. Expansion of DT domains; 7. Efficient mapping of cyber-physical data; and 8. Cloud/Edge computing\footnote{It is worth noticing that literature differentiates Fog, Edge, and Cloud computing~\cite{flammini2021digital}, ie, \textit{Edge} corresponds to field device level, e.g., a smart sensor/actuator whereas \textit{Fog} is a local or a metropolitan area network, for instance, a gateway or local server.
Finally, \textit{Cloud} is a wide area network, e.g., a datacentre in a remote geographical location.}.

\begin{figure}[!htb]
  \centering
  \includegraphics[width=\linewidth]{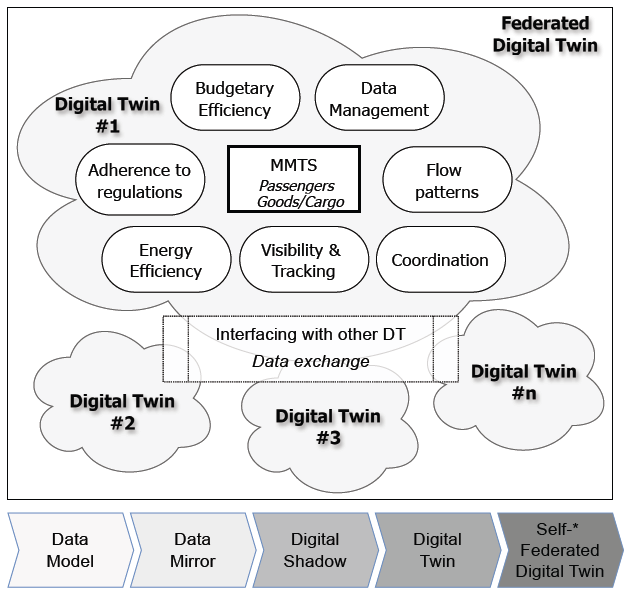}
  \caption{Transitioning from data models to a full-fledged self-* FDT for multinational MMTS.}
  \Description{Boxes representing MMTS functions and interfacing among multiple DT.}
  \label{fig:transitioning}
\end{figure}

We suggest next a way of engaging across multiple stakeholders for establishing the principles for a smooth transition towards a self-* FDT.
Figure ~\ref{fig:transitioning} highlights some of the concerns and objectives of MMTS and DT and allowing multinational organisations to participate in data exchange by understanding the fundamentals of requirements for FDT.

The idea is to allow other infrastructure managers to incorporate other DT into the MMTS provision, by respecting previously accorded data exchange mechanisms. 
Enabling multinational MMTS is not trivial as it involves multiple stakeholders across countries, as well as integrating a plethora of technologies and observing regulatory frameworks.
Key players operating across countries exchanging (near) real-time must establish interoperable ways that makes sense to operators in management settings, filtering out relevant data in a timely fashion.
We envision a hybrid approach with centralised and decentralised operations for smooth collaboration.
It should be centralised for providing a trustworthy current snapshot of all DT, and decentralised when deciding locally without the need to synchronise activities to `higher' coordinating instances.

Authors discuss how to employ data fusion approaches in DT with significant success~\cite{singh2021data,zhang2022digital,minerva2023artificial}.
Liu et al. (2018)~\cite{liu2018role} used for predictive maintenance in aerospace whereas Kaur et al. (2020)~\cite{kaur2020convergence} considered its convergence with IoT and machine learning.
Research focuses on ways of effectively translate multiple flows of information from raw data towards for decision making using sensor-to-sensor, sensor-to-model, and model-to-model fusion approaches.

Next, we highlight key notions in DT/FDT and major challenges.
\begin{itemize}
   \item \textbf{Data management}: Data exchange is key in DT/FDT thus managers must built in their solutions with data management concerns for timely decision making.
   \item \textbf{Interoperability:} Seamless data exchange over multiple DT for advanced analytics.
   \item \textbf{Synchronisation:} Ability of the underlying systems to mirror real-world objects using communication elements that is often assumed to happen in real-time.
   \item \textbf{Adherence to standards:} Need for adherence to multiple standards and multinational data compliance.
   \item \textbf{System Life-Cycle:} Comments on the (overall) Life-Cycle of systems, products, and services. Commissioning and decommissioning objects in physical space and investigating repercussions on the virtual setting.
   \item \textbf{Observability:} Adopted from Software Engineering~\cite{niedermaier2019observability}, where stakeholders aggregate multiple data sources for grasping all significant events in systems and networks.
   \item \textbf{Roadmap and evolution of DT into FDT:} Effective ways of instantiating DT/FDT in real-world settings. Current DT models are highly technical and there is a need for abstracting underlying constructs to pave the way for a FDT deployment.
\end{itemize}

Technological advances in telecommunications will play a significant role in the real-time needs for DT.
Currently, there is infrastructure under use for 5G across developed society, and authors are beginning to investigate 6G networks as well~\cite{alkhateeb2023real}.
External data sources such as environmental, economic, or business might complement the current analysis by offering a broader scope to thoroughly investigate cause-effects.
One must combine these data points when determining what is important when observing how the system and its underlying objects behave over time and by asking relevant details about it.

\section{Conclusion}\label{s:conc}
DT are virtual replicas of physical settings representing a mapping between the real-world setting and its digital counterpart whereas FDT are sets of collaborating DT.
The former have been successfully deployed across many industries and applications, showcasing its strengths and added value by ensuring predictive analytics, simulation, and decision making seamlessly, the latter is nowadays a reasonable promise posing interesting research challenges.

The research question raised in this work was \textit{``how to operationalise effective Federated DT seamlessly, closely reflecting cyber-physical interactions in multinational and multimodal transportation systems?''}
Albeit an interesting topic, available scientific literature on the topic falls short of meaningful research and opens up several opportunities for new advancements in this area.

Collaboration among stakeholders, including governments, transportation companies and technology providers is essential for effective FDT implementation.
Public-private partnerships pool resources and expertise, accelerating innovation and adoption of FDT solutions.
This work aimed to provide a roadmap outlining development stages, milestones and stakeholder roles is vital.
We stress that ensuring scalability and flexibility through a hybrid approach of centralised and decentralised elements provides resilience and supports localised decision-making.
Integrating sustainability goals into FDT helps businesses meet environmental regulations and promote corporate social responsibility.
FDT contributes to efficient resource use and emissions reduction.

One cannot dismiss that integrated approaches with decentralised control with synchronisation requirements will permeate future systems and networks.
For instance, modern transportation systems are intrinsically linked with the energy grid (Smart Grid) and citizen's mobility patterns, and so on.
The DT/FDT provision is well-positioned to stand as the platform for tackling such integration by aggregating and consolidating data produced by a plethora of SoS.
Transitioning from single data models, mirrors (ie, Digital Shadows), then DT, and finally an autonomous, self-operating FDT is not trivial and requires time, investment, and expertise.
The current literature has so far focused on the technical implementation of DT, leaving out the abstraction of the constructs for interoperability.



\printbibliography


\end{document}